\documentclass[runningheads]{llncs}
\usepackage{times}
\usepackage{url}
\usepackage{fancyvrb}
\usepackage{color}
\usepackage{ifthen}
\usepackage{calc}
\begin{document}

\title{Extending the logical update view with transaction support}
\titlerunning{Prolog transactions}

\author{Jan Wielemaker}
\institute{Web and Media group,
	   VU University Amsterdam, \\
	   De Boelelaan 1081a, \\
	   1081 HV Amsterdam,
	   The Netherlands, \\
	   \email{J.Wielemaker@vu.nl}}

\maketitle
\bgroup
\makeatletter
\newcommand{\reffont}{\tt}
\newcommand{\predref}[2]{{\bf #1/#2}}
\newenvironment{itemlist}
  {\itemize
   \renewcommand\makelabel[1]{%
     \hspace\labelwidth
     \llap{\@itemlabel}%
     \hspace\labelsep
     \makebox[\linewidth][l]{\it ##1}%
     \hspace{-\labelsep}}%
     }%
  {\enditemize}
\definecolor{codeboxcolor}{rgb}{0.4,0.4,0.4}
\DefineVerbatimEnvironment%
  {code}{Verbatim}
  {frame=single,
   framerule=0.2pt,
   rulecolor=\color{codeboxcolor},
  }
\newcommand{\const}[1]{{\tt #1}}
\renewcommand{\arg}[1]{\ifmmode\mbox{\em #1}\else{\it #1}\fi}
\newcommand{\secref}[1]{section~\ref{sec:#1}}
\def\term{}
\renewcommand{\term}[2]{%
	\ifthenelse{\equal{\protect}{\protect#2}}{%
	    {\reffont #1}}{%
	    {\reffont #1}({\it #2})}}
\catcode`\^^A=8
\catcode`\_=\active
\def_{\ifmmode\else\_\fi}
\newcommand{\bnfmeta}[1]{\ifmmode{\langle\mbox{\it #1}\rangle}\else$\langle\mbox{\it #1}\rangle$\fi}

\newcommand{\onlinebreak}{}
\def\@nodescription{false}
\newcommand{\defentry}[1]{\definition{#1}}
\renewcommand{\definition}[1]{%
	\onlinebreak%
	\ifthenelse{\equal{\@nodescription}{true}}{%
	    \def\@nodescription{false}%
	    \makebox[-\leftmargin]{\mbox{}}\makebox[\linewidth+\leftmargin-1ex][l]{\bf #1}\\}{%
	    \item[{\makebox[\linewidth+\leftmargin-1ex][l]{#1}}]}}
\def\predatt#1{\hfill{\it\footnotesize[#1]}}
\def\predicate{\@ifnextchar[{\@attpredicate}{\@predicate}}
\def\qpredicate{\@ifnextchar[{\@attqpredicate}{\@qpredicate}}
\def\@predicate#1#2#3{%
	\ifthenelse{\equal{#2}{0}}{%
	    \defentry{#1}}{%
	    \defentry{#1({\it #3})}}%
	\index{#1/#2}\ignorespaces}
\def\@attpredicate[#1]#2#3#4{%
	\ifthenelse{\equal{#3}{0}}{%
	    \defentry{#2\predatt{#1}}}{%
	    \defentry{#2({\it #4})\predatt{#1}}}%
	\index{#2/#3}\ignorespaces}
\def\@qpredicate#1#2#3#4{%
	\ifthenelse{\equal{#3}{0}}{%
	    \defentry{#1:#2}}{%
	    \defentry{#1:#2({\it #4})}}%
	\index{#1/#2}\ignorespaces}
\def\@attqpredicate[#1]#2#3#4#5{%
	\ifthenelse{\equal{#4}{0}}{%
	    \defentry{#2:#3\predatt{#1}}}{%
	    \defentry{#2:#3({\it #5})\predatt{#1}}}%
	\index{#2/#3}\ignorespaces}

\newcommand{\errorterm}[2]{\mbox{\tt%
	\ifthenelse{\equal{}{#2}}{%
	    error(#1, _)}{%
	    error(#1(#2), _)}}}
\newcommand{\termitem}[2]{%
	\ifthenelse{\equal{}{#2}}{%
	    \definition{#1}}{%
	    \definition{#1({\it #2})}}\ignorespaces}
\makeatother

\begin{abstract}
Since the database update view was standardised in the Prolog ISO
standard, the so called logical update view is available in all actively
maintained Prolog systems. While this update view provided a well
defined update semantics and allows for efficient handling of dynamic
code, it does not help in maintaining consistency of the dynamic
database. With the introduction of multiple threads and deployment of
Prolog in continuously running server applications, consistency of the
dynamic database becomes important.

In this article, we propose an extension to the generation-based
implementation of the logical update view that supports transactions.
Generation-based transactions have been implemented according to this
description in the SWI-Prolog RDF store. The aim of this paper is to
motivate transactions, outline an implementation and generate discussion
on the desirable semantics and interface prior to implementation.
\end{abstract}


\section{Introduction}
\label{sec:intro}

Although Prolog can be considered a deductive database system, its
practice with regard to database update semantics is rather poor. Old
systems typically implemented the \textit{immediate update view}, where
changes to the clause-set become immediately visible to all goals on
backtracking. This update view implies that a call to a dynamic
predicate must leave a choice point to anticipate the possibility
that a clause is added that matches the current goal. Quintus
introduced\footnote{\url{http://dtai.cs.kuleuven.be/projects/ALP/newsletter/archive_93_96/net/systems/update.html}}
the notion of the \textit{logical update view}
\cite{DBLP:conf/iclp/LindholmO87}, where the set of visible clauses
for a goal is frozen at the start of the goal, which allows for pruning
the choice-point on a dynamic predicate if no more clauses match the
current goal. Through the ISO standard (section 7.5.4 of ISO/IEC 1321
l-l), the logical update view is adopted by all actively maintained
implementations of the Prolog language.

The logical update view helps to realise efficient programs that depend
on the dynamic database, but does not guarantee consistency of the
database if multiple changes of the database are required to realise a
transition from one consistent state to the next. A typical example is
an application that realises a transfer between two accounts as shown
below.

\begin{code}
transfer(From, To, Amount) :-
        retract(balance(From, FromBalanceStart)),
        retract(balance(To, ToBalanceStart)),
        FromBalance is FromBalanceStart - Amount,
        ToBalance is ToBalanceStart + Amount,
        asserta(balance(From, FromBalance)),
        asserta(balance(To, ToBalance)).
\end{code}

\noindent
Even in a single threaded environment, this code may be subject to
timeouts, (resource-)exceptions and exceptions due to programming errors
that result in an inconsistent database. Many Prolog programs contain a
predicate to clean the dynamic database to avoid the need to restart the
program during development. Still, such a predicate needs to be kept
consistent with the used set of dynamic predicates and provides no
easy solution if the database is not empty at the start.

Obviously, in concurrent applications we need additional measures to
guarantee consistency with concurrent transfer requests and enquiries for
the current balance. Below we give a possible solution based on mutexes.
Another solution is to introduce a \const{bank} thread and realise
transfer as well as enquiries with message passing to the \const{bank}
thread.

\begin{code}
mt_transfer(From, To, Amount) :-
        with_mutex(bank, transfer(From, To, Amount)).

mt_balance(Account, Balance) :-
        with_mutex(bank, balance(Account, Balance)).
\end{code}

\noindent
As we need to serialise update operations in a multi threaded context,
update operations that require significant time to complete seriously
harm concurrency.

With transactions, we can rewrite the above using the code below, where
we do not need any precautions for reading the current
balance.\footnote{Reading multiple values from a single consistent view
still requires a transaction. See \secref{snapshot}.} Consistency is
guaranteed because, as we will explain later, two transactions that
retract the same clause are considered to conflict.

\begin{code}
mt_transfer(From, To, Amount) :-
        transaction(transfer(From, To, Amount), true, [restart(true)]).
\end{code}

\noindent
In the remainder of this article, we first describe related work. Next,
we define the desired semantics of transactions in Prolog, followed by a
description how these semantics can be realised using generations. In
\secref{api}, we propose a concrete set of predicates to make
transactions available to the Prolog programmer. We conclude with
implementation experience in the SWI-Prolog RDF store and a discussion
section.

\section{Related work}

The most comprehensive overview of transactions in relation to Prolog we
found is \cite{Bonner93databaseprogramming}, which introduces
``Transaction logic''. Transaction logic has been implemented as a
prototype for XSB Prolog \cite{Hung96}. This is a much more fundamental
solution in dealing with update semantics than what we propose in this
article. In \cite{Bonner93databaseprogramming}, we also find
descriptions of related work, notably \textit{Dynamic Prolog} and an
extension to Datalog by Naqvi and Krishnamurthy. These systems too
introduce additional logic and are not targeted to deal with
concurrency.

Contrary to these systems, we propose something that is easy to
implement on engines that already provide the logical update view and is
easy to understand and use for a typical Prolog programmer. What we do
learn from these systems is that a backtrackable dynamic database has
promising applications. We not propose to support backtracking
modifications to the dynamic database yet, but our proposal simplifies
later implementation thereof, while the transaction interface may
provide an adequate way to scope backtracking. See \index{transaction/3}\predref{transaction}{3}
described in \secref{api}.

\section{Transaction semantics}

Commonly seen properties of transactions are known by the term
ACID,\footnote{\url{http://en.wikipedia.org/wiki/ACID}} summarised
below. We want to realise all these properties, except for
\textit{Durability}.

\begin{description}
    \item [Atomic]
Either all modifications in the transaction persist or none of
the modifications.
    \item [Consistency]
A successful transaction brings the system from one consistent
state into the next.
    \item [Isolation]
The modifications made inside a transaction are not visible to the
outside world before the transaction is committed.  Concurrent access
always sees a consistent database.
    \item [Durability]
The effects of a committed transaction remain permanently visible.
\end{description}

In addition, code that is executed in the context of a transaction
should behave according to the traditional Prolog (logical view) update
semantics and it must be possible to nest transactions, such that code
that creates a transaction can be called from any context, both outside
and inside a transaction.

An obvious baseline interface for dealing with transactions is to
introduce a meta-predicate \term{transaction}{:Goal}. If \arg{Goal}
succeeds, the transaction is committed. If \arg{Goal} fails or throws an
exception, the transaction is rolled back and \index{transaction/1}\predref{transaction}{1} fails or
re-throws the exception. In our view, \index{transaction/1}\predref{transaction}{1} is logically
equivalent to \index{once/1}\predref{once}{1} (i.e., it prunes possibly remaining choice
points) because database actions are still considered side-effects. Too
much real-world code is intended to be deterministic, but leaves
unwanted choice points. This is also the reason why \index{with_mutex/2}\predref{with_mutex}{2} prunes
choice points.

Note that it is easy to see useful application scenarios for
non-deterministic transactions. For example, generate-and-test
applications that use the database could be implemented using the
skeleton code below.  To support this style of programming, failing
into a transaction should atomically make the changes invisible to
the outside and all changes after the last choice point inside the
transaction must be discarded.  This can be implemented by extending
each choice point with a reference into the change-log maintained
by the current transaction.

\begin{code}
generate_and_test :-
        transaction(generate_world),
        satisfying_world,
        !.
\end{code}

\noindent
Given our proposed once-based semantics, we can still improve
considerably on this use-case compared to traditional Prolog using a
side-effect free generator. This results in the following skeleton:

\begin{code}
generate_and_test :-
        generate_world(World),
        transaction(( assert_world(World),
                      satisfying_world
                    )),
        !.
\end{code}

\noindent
\subsection{Snapshots}
\label{sec:snapshot}

We can exploit the isolation feature of transactions to realise
\textit{snapshots}. A predicate \term{snapshot}{:Goal} executes
\arg{Goal} as \index{once/1}\predref{once}{1} without globally visible affects on the dynamic
database. This feature is a supplement to SWI-Prolog's \textit{thread
local} predicates, predicates that have a different set of clauses in
each thread. Snapshots provide a comfortable primitive for computations
that make temporary use of the dynamic database.  At the same time they
make such code thread-safe as well as safe for failed or incomplete
cleanup due to exceptions or programming errors.

Snapshots also form a natural abstraction to read multiple values from a
consistent state of the dynamic database. For example, the summed
balance of a list of accounts can be computed using the code below. The
snapshot isolation guarantees that the result correctly represents that
summed balance at the time that the snapshot was started.

\begin{code}
summed_balance(Accounts, Sum) :-
        snapshot(maplist(balance, Accounts, Balances)),
        sum_list(Balances, Sum).
\end{code}

\noindent
Note that this sum may be outdated before the isolated goal finishes.
Still, it represents a figure that was true at a particular point in
time, while unprotected execution can compute a value that was never
correct. For example, consider the sequence of events below, where to
summed balance is 10\$ too high.

  \begin{enumerate}
  \item The `summer' fetches the balance of $A$
  \item A concurrent operation transfers 10\$ from $A$ to $B$
  \item The `summer' fetches the balance of $B$
  \end{enumerate}

\section{Generation based transactions}
\label{sec:genupdate}

A common technique used to implement the Prolog logical update view
is to tag each clause with two integers: the generation in which it was
born and the generation in which it died. A new goal saves the current
generation and only considers clauses created before and not died before
its generation. This is clearly described in \cite{Boerger91}. Below, we
outline the steps to add transactions to this picture.

First, we split the generation range into two areas: the low values,
0..G_TBASE (transaction base generation) are used for globally visible
clauses. Generations above G_TBASE are used for generations, where we
split the space further by thread-id. E.g., the generation for the 10th
modification inside a transaction executed by thread~3 is
\mbox{G_TBASE+3*G_TMAX+10}.

\paragraph{Isolation} Isolated behaviour inside a transaction is
achieved by setting the modification generation to the next thread write
generation. Code operating outside a transaction does not see these
modifications because they are time-stamped `in the future'.  Code
operating inside the transaction combines the global view at the start
of the transaction with changes made inside the transaction, i.e.,
changes in the range
\mbox{G_TBASE+\bnfmeta{tid}*G_TMAX..G_TBASE+(\bnfmeta{tid}+1)*G_TMAX}.

\paragraph{Atomic} Committing a transaction renumbers all modifications
to the current global write generation and then increments the
generation. This implies that commit operations must be serialised
(locked). All modifications become atomically visible at the moment that
the global generation is incremented. If a transaction is discarded, all
asserted clauses are made available for garbage collection and the
generation of all retracted clauses is reset to infinity.

\paragraph{Consistency} The above does not provide consistency
guarantees. We add a global consistency check by disallowing multiple
retracts of the same clause. This implies that an attempt to retract an
already retracted clause inside a transaction or while the transaction
commits causes the transaction to be aborted. This constraint ensures
that code that \textit{updates} the database by retracting a value,
computing the new value and asserting this becomes safe. For example,
this deals efficiently with global counters or the balance example from
\secref{intro}. Note that disallowing multiple retracts is also needed
because there is only one placeholder to store the `died generation'.
Similar to relational databases, we can add an integrity constraint,
introducing \term{transaction}{:Goal, :Constraint}, where
\arg{Constraint} is executed while the global commit lock is held.

\paragraph{Nesting} Where we need distinct generation ranges for
concurrently executing transactions, we can use the generation range of
the parent transaction for a nested transaction because execution as
\index{once/1}\predref{once}{1} guarantees strict nesting. Nested transactions merely need to
remember where the nested transaction started. Committing is a no-op,
while discarding is the same as discarding an outer transaction, but
only affecting modifications after the start of the nested transaction.

\paragraph{Implication for visibility rules}

The logic to decide that a clause is visible does not change for queries
outside transactions because manipulations inside transactions are `in
the future'. Inside a transaction, we must exclude globally visible
clauses that have died inside the transaction (i.e., between the
transaction start generation and the current generation) and include
clauses that are created and not yet retracted in the transaction.

\section{Proposed predicates}
\label{sec:api}

We propose to add the following three predicates to Prolog. In the
description below, predicate arguments are prefixed with a \textit{mode
annotation}. The \verb$:$ annotation means that the argument is
module-sensitive, e.g., \arg{:Goal} means that \arg{Goal} is called in
the module that calls the transaction interface predicate. The modes
\verb$+$, \verb$-$ and \verb$?$ specifies that the argument is `input',
`output' and `either input or output'.

\begin{description}
    \predicate{transaction}{1}{:Goal}
Execute \arg{Goal} in a transaction. \arg{Goal} is executed as by
\index{once/1}\predref{once}{1}. Changes to the dynamic database become visible atomically when
\arg{Goal} succeeds. Changes are discarded if \arg{Goal} does not
succeed.

    \predicate{transaction}{2}{:Goal, :Constraint} Run \arg{Goal} as
\index{transaction/1}\predref{transaction}{1}. If \arg{Goal} succeeds, execute \arg{Constraint} while
holding the \const{transaction} mutex (see
\index{with_mutex/2}\predref{with_mutex}{2}\footnote{\url{http://www.swi-prolog.org/pldoc/doc_for?object=with_mutex/2}}).
If \arg{Constraint} succeeds, the transaction is committed. Otherwise,
the transaction is discarded. If \arg{Constraint} fails, throw the error
\errorterm{transaction_error}{constraint, failed}. If \arg{Constraint}
throws an exception, rethrow this exception.

    \predicate{transaction}{3}{:Goal, :Constraint, +Options}
As \index{transaction/3}\predref{transaction}{3}, processing the following options:

    \begin{description}
        \termitem{restart}{+Boolean}
If \const{true}, catch errors that unify with
\errorterm{transaction_error}{_,_} and restart the transaction.
\termitem{id}{Term} Give the transaction an identifier. This identifier
is made available through \index{transaction_property/2}\predref{transaction_property}{2}. There are no
restrictions on the type or instantiation of \arg{Term}.
    \end{description}

    \predicate{snapshot}{1}{:Goal}
Execute \arg{Goal} as \index{once/1}\predref{once}{1}, isolating changes to the dynamic database
and discarding these changes when \arg{Goal} completes, regardless how.

    \predicate{transaction_property}{2}{?Transaction, ?Property}
True when this goal is executing inside a transaction identified by the
opaque ground term \arg{Transaction} and has given \arg{Property}.
Defined properties are:

    \begin{description}
	\termitem{level}{-Level}
Transaction is nested at this level.  The outermost transaction has
level~1.  This property is always present.
	\termitem{modified}{-Boolean}
True if the transaction has modified the dynamic database.
	\termitem{modifications}{-List}
List expresses all modifications executed inside the transaction. Each
element is either a term \term{retract}{Term},\term{asserta}{Term} or
\term{assertz}{Term}. Clauses that are both asserted and erased inside
the transaction are omitted.
	\termitem{id}{?Id}
Transaction has been given the current \arg{Id} using \index{transaction/3}\predref{transaction}{3}.
    \end{description}
\end{description}

If any of these predicate encounters a conflicting retract operation,
the exception \errorterm{transaction_error}{conflict,
PredicateIndicator} is generated.

\section{Implementation results: the SWI-Prolog RDF-DB}

The SWI-Prolog RDF database \cite{DBLP:conf/semweb/WielemakerSW03} is a
dedicated C-based implementation of a single dynamic predicate
\term{rdf}{?Subject, ?Predicate, ?Object}. The dedicated implementation
was introduced to reduce memory usage and improve performance by
exploiting known features of this predicate. For example, all arguments
are ground, and \arg{Subject} and \arg{Predicate} are know to be atoms.
Also, all `clauses' are unit clauses (i.e., there are no rules). Quite
early in the development of the RDF store we added transactions to
provide better consistency and grouping of modifications. These features
were crucial for robustness and `undo' support in the graphical triple
editor Triple20 \cite{DBLP:conf/semweb/WielemakerSW05}. Initially, the
RDF store did not support concurrency. Later, this was added based on
`read/write locks', i.e., multiple readers or one writer may access the
database at any point in time.

With version~3\footnote{Available from
\url{http://www.swi-prolog.org/git/packages/semweb.git}, branch
\texttt{version3}.} of the RDF store, developed last year, we realised
the logical update view also for the external \index{rdf/3}\predref{rdf}{3} predicate and we
implemented transactions following this article. In addition, we
realised concurrent garbage collection of dead triples. The garbage
collector examines the running queries and transactions to find the
oldest active generation and walks the linked lists of the index
hash-tables, removing dead triples from these lists. Actual reclaiming
of the dead triples is left to the Boehm-Demers-Weiser conservative
garbage
collector.\footnote{\url{http://www.hpl.hp.com/personal/Hans_Boehm/gc/}}

\section{Discussion}

We have described an extension to the generation-based logical update
view available in todays Prolog system that realises transactions. There
is no additional memory usage needed for clauses. The engine (each
engine in multi threaded Prolog systems) is required to maintain a stack
of transaction records, where each transaction remembers the global
generation in which is was created and set of affected clauses (either
asserted or retracted). The visibility test of a clause for goals
outside transactions is equal to the test required for realising the
logical update view and requires an additional test of the same cost if
a transaction is in progress.

The described implementation of transactions realises ACI of the ACID
model (atomic, consistency and isolation, but not durability).
Durability can be realised by using a constraint goal that uses
\index{transaction_property/2}\predref{transaction_property}{2} to examine the modifications and write the
modifications to a journal file or external persistent store.

Our implementation has two limitations: (1) goals in a transaction are
executed as \index{once/1}\predref{once}{1} (pruning choice-points) and (2) it is not possible
for multiple transactions to retract the same clause. Supporting
non-deterministic transactions requires additional changes to Prolog
choice points, but transactions already maintain a list of modifications
to realise commit and rollback and \index{transaction/3}\predref{transaction}{3} already provides an
extensible interface to activate this behaviour. Supporting multiple
retracts is possible by using a list to represent the `died' generations
of the clause. This is hardly useful for transactions because multiple
concurrent retracts indicate a conflicting update. However, this
limitation is a serious restriction for snapshots (\secref{snapshot}).

We have implemented this transaction system for the SWI-Prolog RDF
store, where it functions as expected. We believe that transactions will
greatly simplify the implementation of concurrent programs that use the
dynamic database as a shared store. At the same time it eliminates the
need for serialisation of code, improving concurrent performance. In our
experience with Triple20, transactions are useful in single threaded
applications to maintain consistency of the database. Notably,
consistency of the dynamic storage is maintained when an edit operation
fails due to a programming error or an abort initiated from the
debugger. This allows for fixing the problem and retrying the operation
without restarting the application.

We plan to implement the outlined features in SWI-Prolog in the near
future.

\subsection*{Acknowledgements}

This research was partly performed in the context of the COMBINE project
supported by the ONR Global NICOP grant N62909-11-1-7060. This
publication was supported by the Dutch national program COMMIT.

I would like to thank Jacco van Ossenbruggen, Michiel Hildebrand and
Willem van Hage for their feedback in redesigning the transaction
support for the SWI-Prolog RDF store.

\bibliographystyle{plain}
\bibliography{pltrans}

\egroup

\end{document}